# How long do top scientists maintain their stardom? An analysis by region, gender and discipline: evidence from Italy[1]

*Giovanni Abramo*[*]

Laboratory for Studies in Research Evaluation, Institute for System Analysis and Computer Science (IASI-CNR), National Research Council of Italy
Via dei Taurini 19, 00185 Rome, Italy
giovanni.abramo@uniroma2.it

*Ciriaco Andrea D'Angelo*

Department of Engineering and Management, University of Rome "Tor Vergata" and Laboratory for Studies in Research Evaluation, Institute for System Analysis and Computer Science (IASI-CNR)
Via del Politecnico 1, 00133 Rome, Italy
dangelo@dii.uniroma2.it

*Anastasiia Soldatenkova*

Department of Engineering and Management, University of Rome "Tor Vergata"
Via del Politecnico 1, 00133 Rome, Italy
anastasiia.soldatenkova@uniroma2.it

**Abstract**

We investigate the question of how long top scientists retain their stardom. We observe the research performance of all Italian professors in the sciences over three consecutive four-year periods, between 2001 and 2012. The top scientists of the first period are identified on the basis of research productivity, and their performance is then tracked through time. The analyses demonstrate that more than a third of the nation's top scientists maintain this status over the three consecutive periods, with higher shares occurring in the life sciences and lower ones in engineering. Compared to males, females are less likely to maintain top status. There are also regional differences, among which top status is less likely to survive in southern Italy than in the north. Finally we investigate the longevity of unproductive professors, and then check whether the career progress of the top and unproductive scientists is aligned with their respective performances. The results appear to have implications for national policies on academic recruitment and advancement.

[*] *Corresponding author*

# 1. Introduction

One of the distinctive competences of world-class universities is their ability to attract and retain talented professors. They contribute to the prestige of the institutions where they work, attract talented students, receive research grants and funding from companies. The choices faced by faculty hiring committees are then both important and difficult. Especially for younger candidates, this includes the aspect of predicting the likelihood of future scientific success. Recently, decision-makers have increasingly turned to quantitative approaches to inform their judgments, thanks to the rapid development of bibliometrics. Once an appropriate research performance indicator (or set of indicators) is defined, achievements in research can be measured and compared. Scientists with outstanding achievement (typically called "top scientists", or TSs), can be identified as those falling in the top x% out of the total number of their colleagues considering the chosen performance indicator, or as the individuals with performance scores above a certain threshold (for example, the second or the third mean of performance distribution, when analyzed with the *Characteristic Scores and Scales* (*CSS*) technique).

The issues concerning TSs have been of particular interest to scholars in sociology and scientometrics, and they have been studied from different perspectives, among others: in terms of their share of contribution to the overall scientific advancement produced by a research system (Abramo et al., 2013a); in gender analyses (Nowell and Hedges, 1998; Abramo et al., 2009); for the structure of their research collaboration networks (Azoulay et al., 2010); for their roles in the transfer of scientific knowledge to industry (Zucker and Darby, 1997; Zucker et al., 2002; Link et al., 2007).

Given the remarkable contributions in both social and research roles, a number of studies have investigated the possible determinants of the outstanding performance of TSs, over and above their personal merits, such as factors concerning collaboration rates (Lee and Bozeman, 2005), age (Levin and Stephan, 1991), academic origin and affiliation (Long et al., 2009) and incentive systems (Miller et al., 2013). In this line, the literature very often returns to the matter of the Matthew effect (Merton, 1968), which implies that advantage generates further advantage. Merton suggests that eminent scientists will often get more credit than a comparatively unknown researcher, even if their work is similar, meaning that credit would usually be given to researchers who are already famous. Due to this, being a TS at time $t_0$ should increase the likelihood of still being a TS at time $t_1$. Among other authors, Burrell (2003) examines the presence of success-breeds-success phenomenon in the case of citation accumulation. Also, Petersen et. al. (2011; 2014) investigate the presence of cumulative advantage in the careers of scientists and find that up to a certain point, reputation plays a key role in the impact of future publications.

Certainly, the scientist's personal talents must play a decisive role in their rise to the top, or "star" levels. The individual's inculcated talents would not rapidly fade away, and assuming that the Matthew effect holds true, then TSs could maintain their stardom through their entire careers. However, the longevity of TS status would also depend on the dynamics of the individual's external and internal (personal) environment. Within the former, the changes in competition within the field of research, the encounter of barriers to entry, paradigmatic shifts, changes in the direct administrative and working environment, or in availability of resources and collaboration networks, could all affect scientific performance. Within the latter, family changes, changes of interests or moves into managerial positions could affect performance.

With this work, given the range of contrasting factors potentially at play in determining the longevity of the TS status, we attempt to shed light on a first, simple question: how long, in fact, do top scientists maintain their stardom? To the best of our knowledge, the only previous work investigating the longevity of TSs is by Hess and Rothaermel (2012). Their field of observation was the private sector, in particular star scientists in biotech and pharmaceutical industries. The authors employed five-year rolling windows of publication and citation performance in the period 1974-2006. They identified TSs as the scientists having a publication and citation count three standard deviations above the mean for a specific five-year interval. The authors then calculated the number of rolling windows in which the scientist held their star status. The average number of windows for biotech stars was 2.5 (6.5 years), while for those in pharma sector it was 4.1 (8 years). Hess and Rothaermel (2012) conclude that "...the analysis empirically supports the idea that a Matthew effect in science does indeed exist in the biotech and pharmaceutical publishing arenas. In speculation, perhaps the effect is further enhanced by the lack of tenure structure in the corporate setting, serving to keep scientists motivated to publish."

In our own work we observe the longevity of the TSs in an entire higher education system, in particular the top scientists among all Italian professors in all the science disciplines over the period 2001-2012. Because the longevity of a star scientist could in part depend on the field of research, the investigation will examine whether differences in longevity occur at the discipline level, within the national population. We will also check whether the stardom of female TSs is briefer or more prolonged than for males, and if longevity differs across national macro-regions. Finally, we will identify the national population of "unproductive" faculty, and inquire into the longevity of that particular status. In this regard we observe that for legislative and policy purposes, all 96 universities in the Italian national system are research universities, and the responsibilities of all the individual professors include research.

## 2. Data and methods

The population and time-frame for the analysis consists of all Italian professors carrying out research in the so-called hard sciences, over the period 2001-2012. In Italy each professor is classified in one and only one research field. In the hard sciences, there are 205 such fields (named "scientific disciplinary sectors", SDSs[2]), grouped into nine disciplines (named "university disciplinary areas", UDAs[3]). The source for data on the faculty at each university is the database maintained by the Ministry of Education, Universities and Research (MIUR),[4] which indexes the name, gender, academic rank, field (SDS/UDA), and institutional affiliation of all professors in Italian universities, recorded at the close of each year.

The first step is the identification of the TSs, requiring the measurement of research performance for all professors. The bibliometric indicator of performance used is the Fractional Scientific Strength (FSS). The FSS measures the yearly total impact of an individual's research activity over a period of time, adopting the fractional counting method. The advantages of FSS over other per-publication citation indicators, such as the MNCS, are discussed in Abramo and D'Angelo (2016a; 2016b). At present we provide the formula to measure FSS, while referring the reader to Abramo and D'Angelo (2014) for a thorough treatment of the underlying microeconomic theory and all the limits and assumptions embedded in both the definition and the operationalization of the measurement.

---

[2] The complete list is accessible on http://attiministeriali.miur.it/UserFiles/115.htm, last accessed November 14, 2016.
[3] Mathematics and computer sciences, Physics, Chemistry, Earth sciences, Biology, Medicine, Agricultural and veterinary sciences, Civil engineering, Industrial and information engineering.
[4] http://cercauniversita.cineca.it/php5/docenti/cerca.php, last accessed November 14, 2016.

$$FSS = \frac{1}{t}\sum_{i=1}^{N}\frac{c_i}{\bar{c}}f_i$$

[1]

Where:
$t$ = number of years of work in the period under observation
$N$ = number of publications in the period under observation
$c_i$ = citations received by publication $i$
$\bar{c}$ = average of distribution of citations received for all cited publications[5] in same year and subject category of publication $i$
$f_i$ = fractional contribution of professor to publication $i$.

The fractional contribution equals the inverse of the number of authors in those fields where the practice is to place the authors in simple alphabetical order but assumes different weights in other cases. For the life sciences, widespread practice in Italy is for the authors to indicate the various contributions to the published research by the order of the names in the byline. For the life science SDSs, we give different weights to each co-author according to their position in the list of authors and the character of the co-authorship (intra-mural or extra-mural) (Abramo et al., 2013b).[6]

Based on the value of FSS, expressed on a percentile scale of 0-100 (worst to best), we obtain a ranking list of all professors for each SDS. In our analysis we investigate two subsets: TSs as those that place from the 90 percentile up[7], and unproductive professors (UNs) as those with nil FSS.

The bibliometric dataset used to measure FSS is extracted from the Observatory of Public Research (ORP), a database developed by the authors and derived under license from the Thomson-Reuters Italian National Citation Report, an extract of the WoS. Beginning from the raw data of the WoS and applying a complex algorithm for disambiguation of the true identity of the authors and reconciliation of their institutional affiliations, each publication is attributed to the university scientist(s) that produced it, with a harmonic average of precision and recall (F-measure) equal to 97% (for details see D'Angelo et al., 2011). For each publication, the bibliometric dataset thus provides:
- the complete list of all coauthors;

---

[5] Abramo et al. (2012a) demonstrated that the average of the distribution of citations received for all cited publications of the same year and subject category is the most effective scaling factor.

[6] It must be noted that different fractional counting across disciplines does not cause any bias, because the top 10% scientists are extracted from each field. To exemplify, if we did not weight the authors' contribution in Cardiology, the top 10% scientists in cardiology might change, but all the remaining top scientists (from the other fields) would be exactly the same.

[7] In order to check the consistency of the results, we adopt also another definition of TS, as the one whose performance falls above the mean of the subpopulation above the first mean of the overall population in their SDS, by the CSS technique (Glänzel and Schubert, 1988).

- the complete list of all their addresses;
- a sub-list of only the academic authors, with their SDS/UDA and university affiliations.

The observation period of production is 2001-2012, while citations are counted for all publications at a later date (31/05/2015). The citation window is broad enough to ensure robust performance scores for even the latest publications, of 2012 (Abramo et al., 2011).

For examination of the longevity of top-ranked scientists, we split the timeframe into three four-year consecutive periods. In each period we identify the TSs and UNs among the researchers on staff for at least three out of the four years. (With a shorter period of observation the performance scores would be less robust. See Abramo et al., 2012b). Our question is how many of the professors identified as TSs and UNs in the first period remain as such in the subsequent periods. Apart from the analysis of the overall population, the examination will also be carried out also at the gender, discipline and regional levels. The regional location of each TS is assigned on the basis of the university of their employment at the end of the first period (2004).

## 3. Results

We begin the analysis of TS longevity at the aggregate level and continue to the gender, territory and discipline levels. We then replicate the analyses for the UNs.

From the population of Italian professors in the sciences on staff for at least three years in the period 2001-2004 (33168), we extract the top 10% in each SDS (3407), excluding those who were not on staff in the following periods (2005-2008; and 2009-2012). For this first period ("A", 2001-2004), the dataset thus composed consists of 2883 TSs. We then extract the TSs from the populations in the subsequent periods, in similar manner. Finally we check how many 2001-2004 TSs maintain their stardom over the subsequent two periods. Figure 1 presents the Euler diagram for the TSs longevity. The inclusive circle A represents the 2883 TSs in the period 2001-2004, who remained on staff (whether or not as TSs) over the full time-span examined (2001-2012). The second largest circle marked as A∩B represents the 1572 (55%) professors who maintained their top position in period B (2005-2009). The third circle marked as A∩C consists of the 1196 (41%) professors who were TSs in both periods A and C (2009-2012).[8] The intersection of the three periods A∩B∩C represents the 1004 (35%) professors who hold the status of TS through all three periods.

The analysis of gender differences in longevity shows that 16% of TSs in the first period are female, but only 13% maintain stardom over all three periods (Table 1). The concentration indices relative to the TSs across all three periods are then 0.81 for females and 1.04 for males. Of the 2001-2004 TSs, 36% of the men maintain their stardom until 2012, compared to 28% of the women. Thus it appears that the female scientists are less likely to maintain their stardom.

---

[8] Concerning the intersections of two periods, we repeated the analyses but relaxing the constraint that the TSs must be on staff from three periods to two periods. Under this changed condition, the share of those who maintained their stardom for two periods resulted exactly the same as in Table 1.

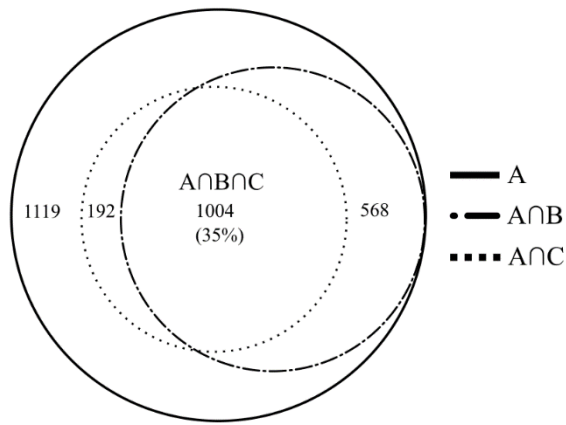

*Figure 1: Euler diagram for TS longevity: circle A represents the set of top 10% scientists in period A (2001-2004); B those in 2005-2008; C those in 2009-2012*

*Table 1: Longevity of top 10% scientists by gender*

| Gender | A | A∩B∩C | A∩B∩C/A | Concentration index |
|---|---|---|---|---|
| M | 2422 (84%) | 873 (87%) | 36% | 1.04 |
| F | 461 (16%) | 131 (13%) | 28% | 0.81 |

*Legend: A represents the set of top scientists in period A (2001-2004); B those in 2005-2008; C those in 2009-2012*

To assess variations in the longevity of TSs across disciplines we conduct the analysis at UDA level. Table 2 reports, for each UDA, the number of TSs in period A and those among them who maintained their top status in the subsequent four-year periods (B, C). The percentages in brackets represent the share of remaining TSs relative to the number in the first period. Our main concern is the intersection A∩B∩C, meaning the scientists who maintain their stardom over all three periods. The minimum share of 20% is observed in Civil engineering (UDA 8), while the maximum of 45% in Biology (UDA 5) followed by Medicine (42%). If a Matthew effect is at work, it seems it is more effective in the life sciences and less in engineering.

*Table 2: Longevity of top 10% scientists at UDA level*

| UDA | A | A∩B | A∩C | A∩B∩C |
|---|---|---|---|---|
| 1 - Mathematics and computer science | 280 | 139 (50%) | 103 (37%) | 77 (28%) |
| 2 - Physics | 214 | 122 (57%) | 75 (35%) | 65 (30%) |
| 3 - Chemistry | 247 | 145 (59%) | 111 (45%) | 99 (40%) |
| 4 - Earth sciences | 101 | 44 (44%) | 30 (30%) | 23 (23%) |
| 5 - Biology | 394 | 255 (65%) | 206 (52%) | 176 (45%) |
| 6 - Medicine | 843 | 498 (59%) | 405 (48%) | 351 (42%) |
| 7 - Agricultural and veterinary sciences | 265 | 121 (46%) | 92 (35%) | 70 (26%) |
| 8 - Civil engineering | 127 | 47 (37%) | 36 (28%) | 26 (20%) |
| 9 - Industrial and information engineering | 412 | 201 (49%) | 138 (33%) | 117 (28%) |

*Legend: A represents the set of top scientists in period A (2001-2004); B those in 2005-2008; C those in 2009-2012*

We now check the consistency of the above results, carrying out the same analysis for TSs defined as those whose performance is above the second mean (TSs$_{\mu2}$), when the performance is analyzed by the CSS technique. Table 3 shows that results are aligned to the above ones.

*Table 3: Longevity of TS$_{\mu2}$ at UDA level*

| UDA | A | A∩B | A∩C | A∩B∩C |
|---|---|---|---|---|
| 1 - Mathematics and computer science | 231 | 101(44%) | 95(41%) | 65(28%) |
| 2 - Physics | 162 | 79(49%) | 66(41%) | 58(36%) |
| 3 - Chemistry | 213 | 122(57%) | 101(47%) | 84(39%) |
| 4 - Earth sciences | 83 | 48(58%) | 31(37%) | 25(30%) |
| 5 - Biology | 265 | 182(69%) | 152(57%) | 127(48%) |
| 6 - Medicine | 612 | 374(61%) | 307(50%) | 252(41%) |
| 7 - Agricultural and veterinary sciences | 208 | 107(51%) | 77(37%) | 54(26%) |
| 8 - Civil engineering | 92 | 50(54%) | 36(39%) | 24(26%) |
| 9 - Industrial and information engineering | 312 | 174(56%) | 123(39%) | 97(31%) |

*Legend: A represents the TSs$_{\mu2}$ in period A (2001-2004); B those in 2005-2008; C those in 2009-2012*

Finally we investigate the differences in longevity at the regional level. The Italian territory is divided into 20 administrative regions, grouped for various considerations as three macro-regions: north, center, and south. The south has a history of slower industrial and economic development than the north (Daniele and Malanima, 2011; SVIMEZ, 2015; ISTAT, 2015). Given that the characteristics of the higher education system also show a north-south divide (Viesti, 2015; Abramo et al., 2016), we are interested in assessing whether such a pattern extends to the longevity of TSs.

Table 4 presents the distribution of TSs by macro-region. 52% of the 2001-2004 TSs are based in universities located in the north, compared to 25% in central Italy and 23% in the south. Out of the total TSs who maintain their stardom for three consecutive periods, 55% are from the north, 26% from the center, and 18% from the south. The concentration indexes then are respectively 1.07 in the north, 1.04 in the center, and 0.79 in the south. Not only are there fewer TSs in the south, their longevity is also less compared to the other macro-regions. Looking at the 2001-2004 TSs, 37% of those from the north and 36% from the center maintain their stardom over three consecutive periods, while only 28% from the south experience this success.

*Table 4: Longevity of top 10% scientists by macro-region*

| Macro-region | A | A∩B∩C | A∩B∩C/A | Concentration index |
|---|---|---|---|---|
| North | 1489 (52%) | 555 (55%) | 37% | 1.07 |
| Center | 733 (25%) | 266 (26%) | 36% | 1.04 |
| South | 661 (23%) | 183 (18%) | 28% | 0.79 |
| Entire nation | 2883 | 1004 | 35% | |

*Legend: A represents the set of top scientists in period A (2001-2004); B those in 2005-2008; C those in 2009-2012*

We now replicate the analysis for the unproductive scientists, or UNs.[9] The occurrence of UN academics should be quite exceptional, and particularly given a national policy requiring research, and one would hope for nil longevity of any instances. However, as Abramo et al. (2013a) have shown, the case of unproductive professors is all too common in Italy. Our question here is about the longevity of such faculty.

Looking at the population of 33168 sciences professors in 2001-2004, 8217 (24.8%) resulted as UNs, which is extraordinary high. Of these, 4703 remained as university faculty over all three periods (19.4% of the corresponding population of professors). As shown in Figure 2, of the 4703 on staff for the entire window, 2517 (54%) remained UNs in the second period (A∩B), and 1680 (36%) were UNs through the third period (A∩B∩C). We observe that the longevity of UNs is similar to TSs.

---

[9] We did not conduct the UN analysis at discipline level, since the differences in shares of UNs are heavily affected by WoS coverage and by publication behaviors unique to the disciplines (Abramo and D'Angelo, 2015).

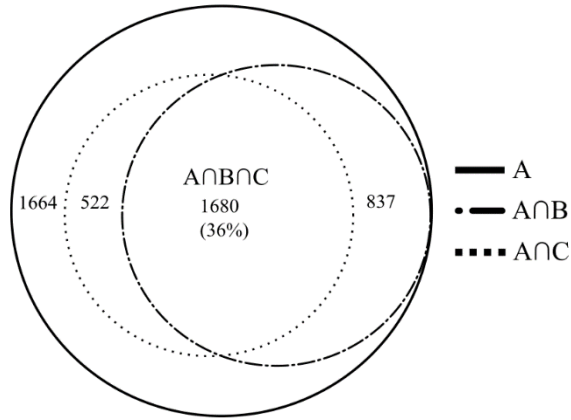

*Figure 2: Euler diagram for unproductive professors' longevity: A represents the set of unproductive professors in period A (2001-2004); B those in 2005-2008; C those in 2009-2012*

The analysis of gender differences demonstrates that female professors make up 26% of the UNs in the first period, and 28% of the unproductive population across all three periods (Table 5). The index of concentration for those remaining unproductive across all periods is 1.07 for females and 0.98 for males. Of the 2001-2004 UNs, 35% of the men and 38% of the women remain unproductive throughout.

The distribution of UNs among the macro-regions is shown in Table 6. In the 2001-2004 period, 37% of UNs were employed by northern universities, 26% in central Italy and 38% in the south. Among those who remain unproductive through all three periods, 39% are in the north, 24% from the center, and 36% in the south. Thus, the concentration indexes of the longest-term UNs are respectively 1.07 in the north, 0.92 in the center, and 0.99 in the south. It appears then that compared to those in the south, the UNs in the north are marginally more capable of maintaining their dubious status.

*Table 5: Longevity of unproductive professors by gender*

| Gender | A | A∩B∩C | A∩B∩C/A | Concentration index |
|---|---|---|---|---|
| M | 3486 (74%) | 1216 (72%) | 35% | 0.98 |
| F | 1217 (26%) | 464 (28%) | 38% | 1.07 |

*Legend: A represents the set of unproductive professors in period A (2001-2004); B those in 2005-2008; C those in 2009-2012*

*Table 6: Longevity of unproductive professors by macro-region*

| Macro-region | A | A∩B∩C | A∩B∩C/A | Concentr. index |
|---|---|---|---|---|
| North | 1724 (37%) | 656 (39%) | 38% | 1.07 |
| Center | 1213 (26%) | 397 (24%) | 33% | 0.92 |
| South | 1766 (38%) | 627 (36%) | 36% | 0.99 |
| Italy | 4703 | 1680 | 36% | |

*Legend: A represents the set of unproductive professors in period A (2001-2004); B those in 2005-2008; C those in 2009-2012*

Given that the competitions for Italian academic recruitment and career advancement are notoriously affected by favoritism, and far from meritocratic (Perotti, 2008; Gerosa, 2011; Abramo et al., 2015), we wished to delve further into the career progress of both the TSs and UNs. In a meritocratic system one would expect that the "three-period" TSs employed in the lower academic ranks would have excellent possibilities of promotion to higher ranks, while no UNs would experience advancement. In fact we find that 39 (24%) of the 165 three-period TS assistant professors were never promoted, while 90 (11%) three-period UN assistant professors were advanced to associate professor. Similarly, 121 (41%) of the three-period TS associate professors were never promoted, while 60 (12%) of three-period UN associates professors did advance to full professorship.

In Italy, the mobility of professors among universities is very low. Only 35 (3.5%) three-period TSs and 11 (0.65%) UNs move from one macro-region to another over the twelve-year period. The net balance of TSs in-outflow is positive in the north and negative for the other two macro-regions. The opposite is true for UNs. Although the numbers are very small, they suggest that, among other factors, the universities in the north may have a superior ability to attract TSs from other macro-regions, while getting rid of their UNs.

### 4. Conclusions

A substantial number of studies have examined the different attributes of top scientists, including inquiring into the factors that could contribute to or detract from achieving and retaining such status. However their longevity, once arrived, has not received significant attention. The current study responds to this basic question. The findings reveal that over 12 years, 35% of top scientists retain their star status for three consecutive four-year periods, and 55% for two periods. The contribution of the Matthew effect to this staying power is difficult to estimate, however if it is at work, it seems more pronounced in the life sciences than in engineering. The results show that female TSs are less successful in maintaining their stardom than males, as could be expected given the role of women in Italian families, especially if with children. There are also regional differences in staying power: the TSs of southern Italy are more likely to lose their top status than those from the center, and even more so than those in the north. This result is aligned with the lower individual and institutional research productivity in the south (Abramo et al. 2016).

It must be noted that under the remuneration policy for Italian academics, starting salaries and their increments are not linked to merit, thus failing to provide an important motivation for improving research productivity, and in particular to maintain stardom. In this same national system, it is not surprising that the already high share of unproductive scientists seems to enjoy comfortable longevity, showing proportions very similar to those of the TSs, to the dismay of any taxpayers. Differently from the female TSs, their gender counterparts among UNs slightly exceed the males in maintaining the laurels of nil productivity. While the analysis revealed that the TSs based in the north have more staying power than those in south, the same regional pattern also holds true for the UNs.

A further result should be of particular interest to Italian policy makers (but not only). In a final analysis we examined the career progress of both the longest-lasting TSs and UNs (at least 12-year duration). We found that remarkable numbers of TSs received no promotion throughout the entire period, and that this was unhappily mirrored by equally remarkable figures for UNs that were awarded advancement. Although scientometric analyses should be regarded as supportive to peer-review decision making, and exceptions can always occur, the results seem to provide in general additional evidence of the widespread phenomenon of favoritism and discrimination in Italian academic public competitions for recruitment, as shown by a number of previous studies and the recent warning by the President of the Italian anti corruption authority.[10]

---

[10] http://www.ansa.it/sito/notizie/topnews/2016/09/23/cantone-allarme-corruzione-universita_687f391c-802c-478b-b9e4-6d4f3c1b8f0d.html, last accessed November 14, 2016.